\documentclass{PoS}
\graphicspath{{./fig/}}
\usepackage{latexsym,graphicx,amssymb,amsmath,mathrsfs}
\usepackage{epstopdf}
\usepackage[normalem]{ulem}

\renewcommand\sout{\bgroup \color{red} \ULdepth=-.5ex \ULset}
\newcommand{\comment}[1]{}

\title{The path optimization for the sign problem of
       low dimensional QCD\footnote{Report No.: KUNS-2780, YITP-19-110}}

\ShortTitle{The path optimization for the sign problem of
            low dimensional QCD}

\author{\speaker{Yuto Mori}\\
        Department of Physics, Faculty of Science,
        Kyoto University,
Kyoto 606-8502, Japan\\
        E-mail: \email{mori.yuto.47z@st.kyoto-u.ac.jp}}

\author{Kouji Kashiwa\\
       Fukuoka Institute of Technology, Wajiro,
       Fukuoka 811-0295, Japan\\
       E-mail: \email{kashiwa@fit.ac.jp}}

\author{Akira Ohnishi\\
       Yukawa Institute for Theoretical Physics,
       Kyoto University, Kyoto 606-8502, Japan\\
       E-mail: \email{ohnishi@yukawa.kyoto-u.ac.jp}}

\abstract{
The path optimization has been proposed to weaken the sign problem
which appears in some field theories such as finite density QCD.
In this method, we optimize the integration path in complex plain to enhance
the average phase factor.
In this study, we discuss the application of this method to
low dimensional QCD as a first step of finite density QCD.
}

\FullConference{37th International Symposium on Lattice Field Theory - Lattice2019\\
		16-22 June 2019\\
		Wuhan, China}

\begin{document}
\section{Introduction}
The sign problem in the path integral is a numerical difficulty
that appears in system with complex actions.
Complex action causes rapid oscillation and
serious cancellation 
 of the Boltzmann weight in calculating the partition function,
and this cancellation makes the numerical accuracy worse.
In lattice QCD at finite chemical potential $\mu$,
Fermion determinant satisfies
$(\mathrm{det}~D(\mu))^* = \mathrm{det}~D(-\mu^*)$,
and generally takes a complex value.
So the sign problem is one of the most important issues because
this makes it difficult to investigate finite density QCD from the
first principles.

Recently,
in order to avoid or evade the sign problem,
some complexified variable methods
such as
the complex Langevin method
~\cite{Parisi:1984cs,Aarts:2009uq,Nagata:2016vkn},
the Lefschetz thimble method
~\cite{Witten:2010cx,Cristoforetti:2012su,Fujii:2013sra,Alexandru:2015sua}
and the path optimization method
~\cite{Mori:2017pne,Mori:2017nwj,Kashiwa:2018vxr,Kashiwa:2019lkv}
have been proposed.
In the Lefschetz thimble method,
one complexifies the dynamical variables
and give the integration path (manifold) by solving the flow equation
from the fixed point or deforming original real path by the flow equation.

On the other hand, in the path optimization method,
the integration path is given by minimization of the cost function,
which represents the seriousness of the sign problem.
This method has been applied to
a one dimensional integral~\cite{Mori:2017pne},
a finite density complex scalar field theory~\cite{Mori:2017nwj},
and the Polyakov loop extended Nambu-Jona-Lasinio model
with and without vector type interaction
~\cite{Kashiwa:2018vxr,Kashiwa:2019lkv}.
In these cases, the average phase factor (APF) is enhanced significantly.

In this proceedings,
we apply the path optimization method
to 0+1D QCD~\cite{Mori:2019tux} and 1+1D QCD at finite chemical potentials
as a first step towards the finite density QCD on four dimensional lattice.
It should be noted that the low dimensional
QCD at finite densities have been studied
by using the Complex Langevin method
(0+1D and 1+1D)~\cite{Aarts:2010gr}
 and by using Lefschetz thimble method~\cite{DiRenzo:2017igr}.

\section{Path Optimization method}

We consider 
the partition function given by the integral of the complex Boltzmann weight
on the complexified integration path $z(x)$,
\begin{align}
  \mathcal{Z} = \int_{\mathcal{C}_\mathbb{R}} d^Nx\, \exp(-S(x)) 
  = \int_{\mathcal{C}_\mathbb{C}} d^Nz\, \exp(-S(z))
  = \int_{\mathcal{C}_\mathbb{R}} d^Nx\, J(z) \exp(-S(z(x))),
\end{align}
where $J(z)=\det (\partial z/\partial x)$ is the Jacobian.
Here we note that the integral of analytic function is independent
of integration path from the Cauchy integral theorem.

To obtain the optimized integration path,
we first give the integration path by a trial function
containing variational parameters,
and next we evaluate the cost function
which represents the degree of the sign problem. We adopt the following
cost function,
\begin{align}
  \mathcal{F}[z(x)]
  &= \int_{\mathcal{C}_\mathbb{R}} d^Nx \left|J(z) \exp(-S(z(x)))\right|
  - |\mathcal{Z}| \\
  &= |\mathcal{Z}| \{|\langle e^{i\theta} \rangle|^{-1}-1\} ,
\end{align}
where $\theta$ is the phase of the statistical weight
$\theta=\arg (J(z) \exp(-S(z)))$ and
$\langle e^{i\theta} \rangle$ is APF.
In the third step, we optimize the variational parameters to 
reduce the cost function.
Hence we can increase APF which determines 
the precision in phase reweighting method.
Finally, we generate field configurations
by the Monte-Carlo method on the optimized path.

We use the neural network
to represent the integration path  theories 
with many degree of freedom. The neural network
is a powerful tool to represent any functions of many inputs.
Neural network with mono-hidden layer is composed of
the input layer, hidden layer, and output layer.
The variables given to a layer are passed to the next layer
after the linear and non-linear transformation,
\begin{align}
  a_i = g(W_{ij}^{(1)} x_j + b_i^{(1)}),~~
  y_i = \alpha_i g(W_{ij}^{(2)}a_j +b_i^{(2)}), \label{Eq:complex_scalar}
\end{align}
where $x_i$ and $y_i$ are input and output variables respectively,
$g(x)$ is called activation function and we use $g(x) = \tanh (x)$,
and $W, b, \alpha$ are the variational parameters.
By using these transformation, any function is known to be represented
in the large number limit of hidden layer units~\cite{cybenko1989approximation}.

Inputs of the neural network are the real integral variables,
and it is natural to take variables
before the complexification. For example, in the scalar field theory,
we input the real part of variables $x_i$,
and obtain the imaginary part from outputs of the neural network $y_i$.
For the $\mathrm{SU}(3)$ gauge theory,
an SU(3) link variable $U$ is complexified to an SL(3) matrix $\mathcal{U}$.
So we input the matrix elements of $U$
and obtain the complexified link variables,
$\mathcal{U} \in \mathrm{SL}(3, \mathbb{C})$ as follows,
\begin{align}
  \mathcal{U}(U) = U \prod_{i=1}^8
    e^{\lambda_i y_i(\mathrm{Re}~U,\mathrm{Im}~U)}, \label{Eq:complex_gauge}
\end{align}
where $\lambda_i$ are the generators of $\mathrm{SU}(3)$.

\section{0+1D QCD}

We here discuss 0+1D lattice QCD based on the results in Ref.~\cite{Mori:2019tux}.
The action of 0+1D lattice QCD with one species of staggered fermion
 consists of temporal fermion hopping terms and mass terms,
\begin{align}
 S &= \frac{1}{2} \sum^{N_\tau}_{\tau=1}
      (\overline{\chi}_\tau e^{\mu} U_\tau \chi_{\tau+1} -
      \overline{\chi}_{\tau+1} e^{-\mu} U^{-1}_\tau \chi_\tau)
      + m \sum_\tau \overline{\chi}_\tau\chi_\tau,
\end{align}
where $N_\tau$ is lattice size,
$m$ and $\mu$ are mass and chemical potential respectively.
The partition function is obtained as
\begin{align}
  \mathcal{Z} &= \int \mathcal{D}U \det(D(U))
  = \int dU \det \left[ X_{N_\tau} + (-1)^{N_\tau} e^{\mu/T}U +
    e^{-\mu/T}U^{-1} \right] \\
  X_{N_\tau} &= 2\cosh (E/T),~
  E = \mathrm{arcsinh}\,m,~U=U_1U_2\cdots U_{N_\tau},~T = 1/N_\tau.
\end{align}
This theory can be reduced to a one-link variable problem
by using the gauge transformation.
It should be noted that no plaquette terms appear
because there is no spatial direction.
Fermion determinant becomes complex at finite chemical potential.

We can further reduce the number of variables from
eight in the $\mathrm{SU}(3)$ link variable
to two real variables $x_i \in [0,2\pi]~(i=1,2)$
by using the diagonalized gauge fixing,
$U=\mathrm{diag}(e^{ix_1},e^{ix_2},e^{-i(x_1+x_2)})$.
So we demonstrate the two ways of complexifying link variables.
In the first way,
we complexify the variables after diagonalized gauge fixing,
we complexify $x_i$ by Eq.~(\ref{Eq:complex_scalar}).
In the second method,
we complexify the link variable without diagonalized gauge fixing,
link variable is complexified by Eq.~(\ref{Eq:complex_gauge}).

\begin{figure}[bthp]
\begin{center}
\includegraphics[trim=40 0 40 0,width=0.33\textwidth]{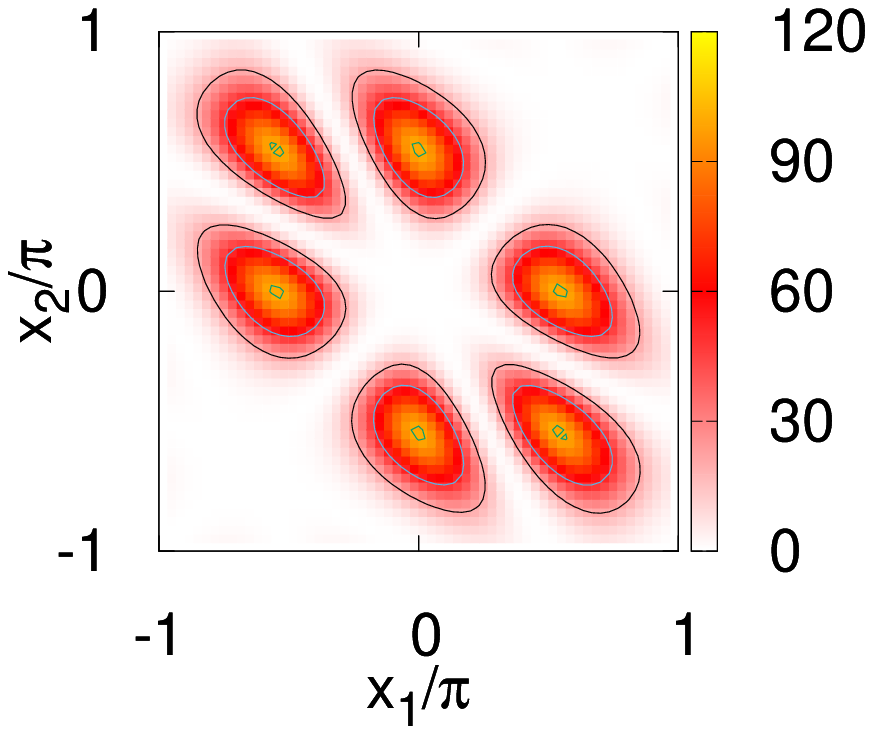}%
\includegraphics[trim=40 0 40 0,width=0.33\textwidth]{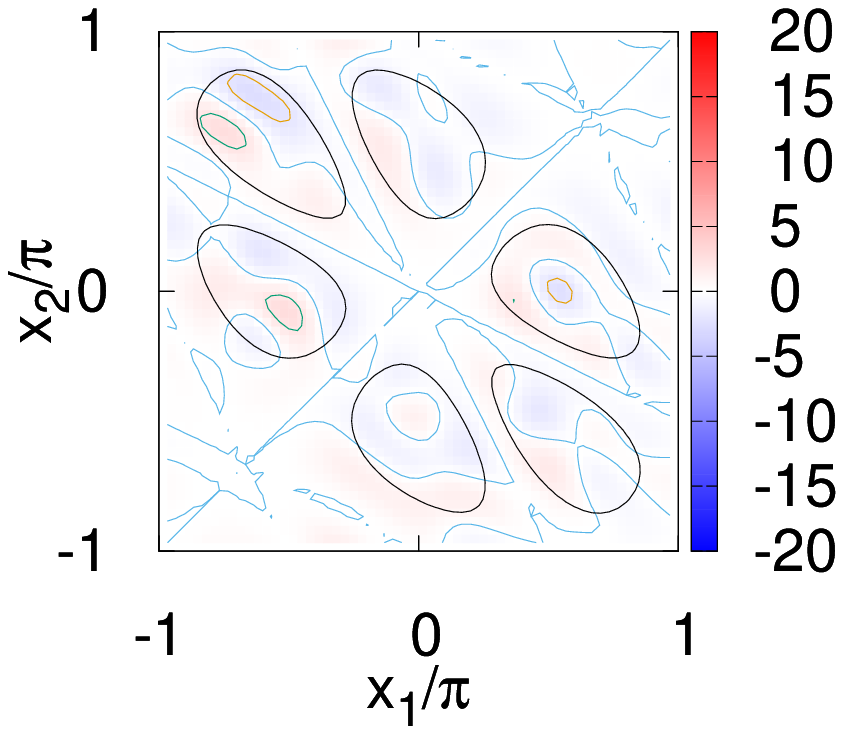}%
\includegraphics[trim=40 0 40 0,width=0.33\textwidth]{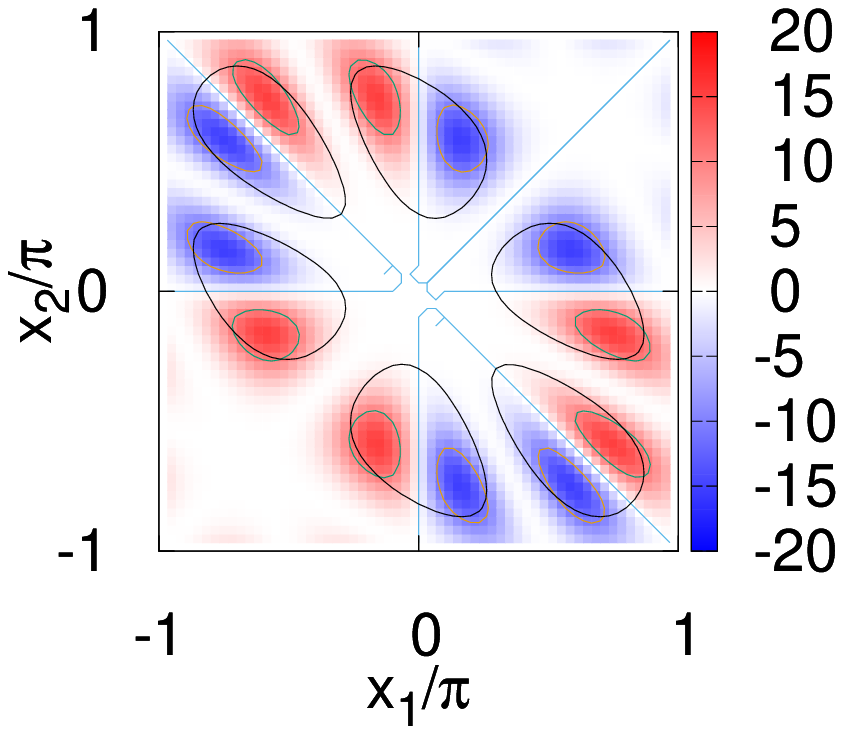}%
\end{center}
\caption{
Absolute value of the statistical weight distribution $|JW|$ with
the  optimization (left),
and imaginary part of the statistical weight distribution
$\mathrm{Im}(JW)$ with (middle) and without (right) the optimization.
Black curves show the contour of $|JW|=20$.
The results are calculated at $\mu/T=1.0$.}
\label{Fig:2D_JW}
\end{figure}

Let us first discuss the results
of complexification after diagonalized gauge fixing.
We optimize the function $z(x)$ by the gradient descent method
to minimize the cost function.
Figure \ref{Fig:2D_JW} shows the statistical weight,
the Jacobian times the Boltzmann factor $Je^{-S}$, as functions
of $x_1$ and $x_2$ with and without the optimization~\cite{Mori:2019tux}.
We find that the there are six separated regions
having large statistical weights, which will be discussed later.
The imaginary part of the statistical weight is suppressed
by the optimization, as seen by the white color showing
$\mathrm{Im}\,(JW) \simeq 0$ in the middle panel.

\begin{figure}[bthp]
\begin{center}
\includegraphics[width=0.33\textwidth]{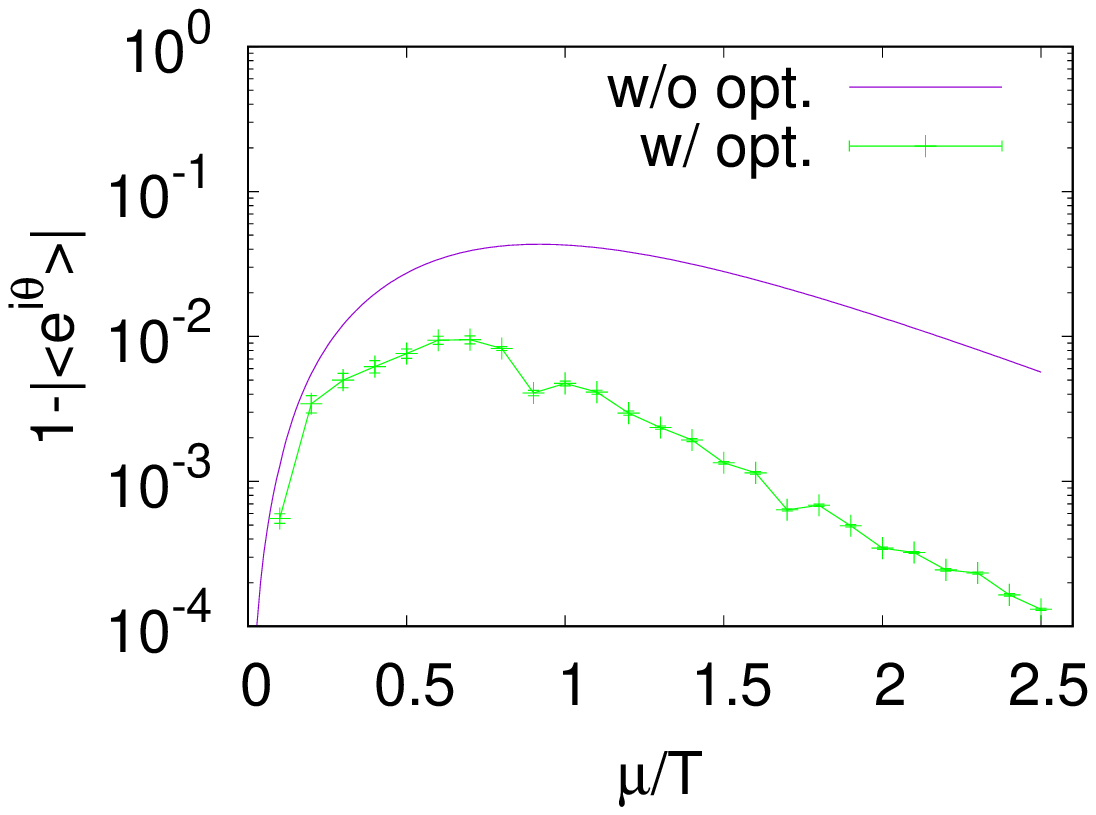}%
\includegraphics[width=0.33\textwidth]{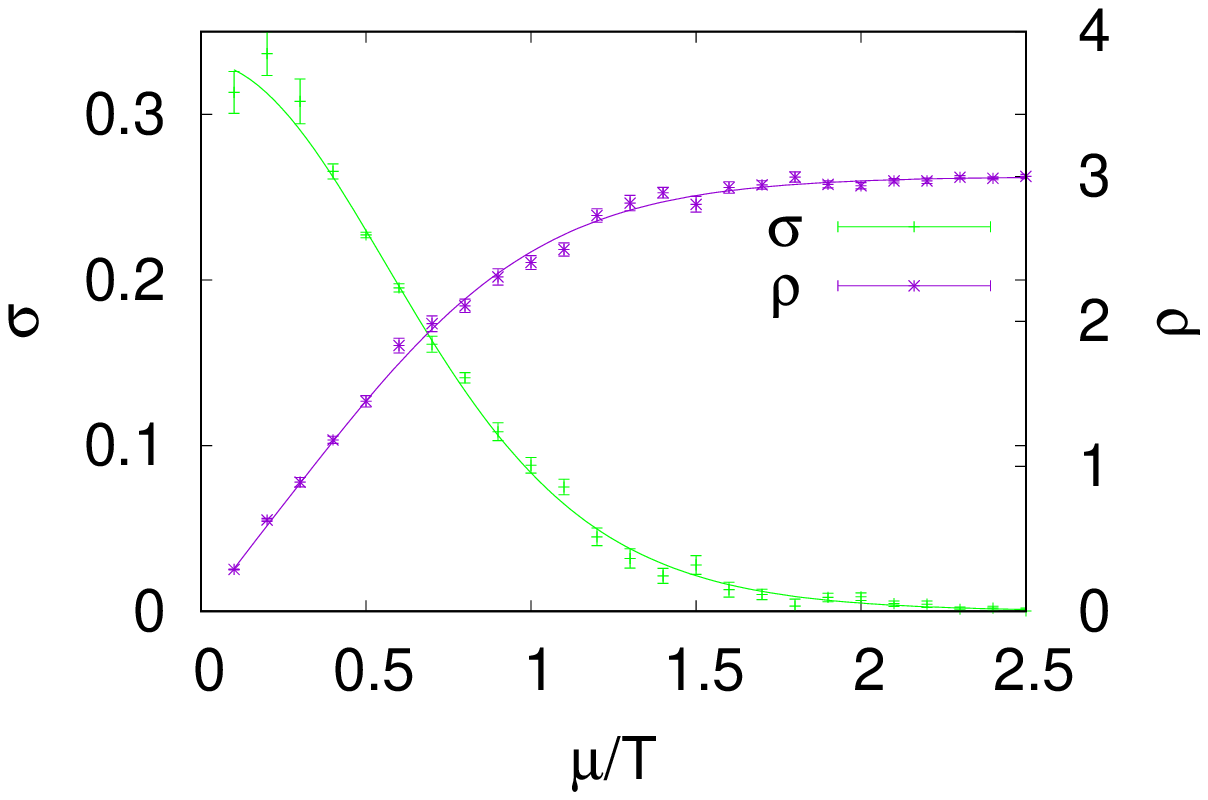}%
\includegraphics[width=0.33\textwidth]{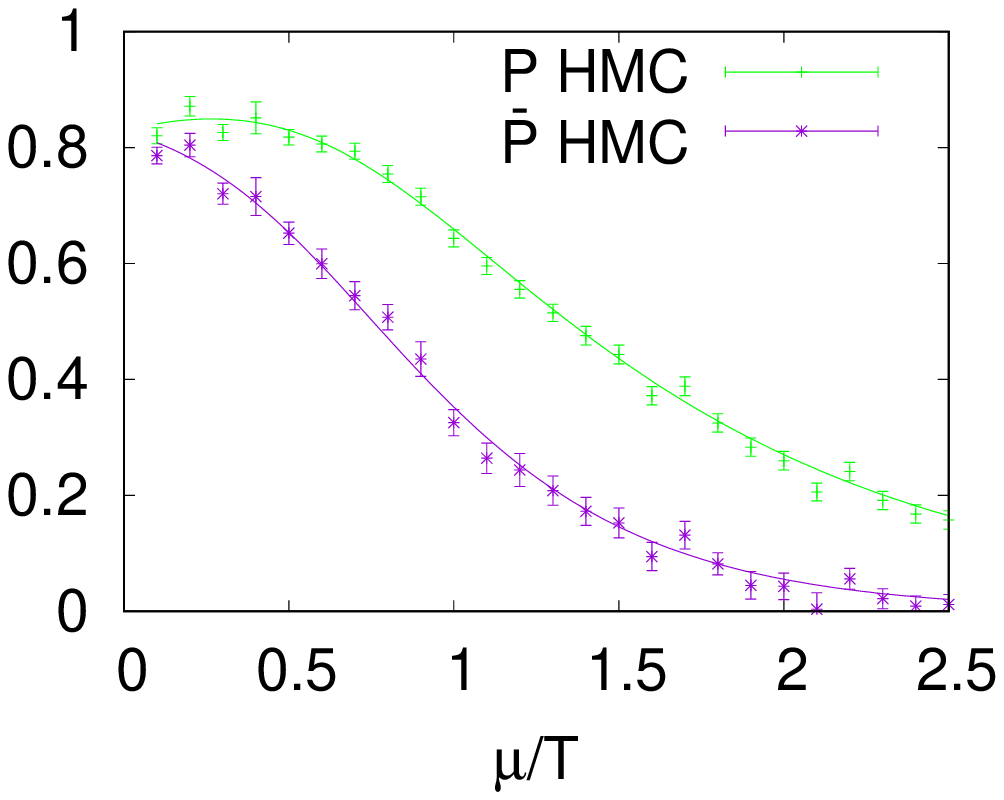}%
\end{center}
\caption{
The average phase factor (left),
the expectation value of quark condensate and quark density (middle),
and that of Polyakov loop (right).
}
\label{Fig:expect}
\end{figure}

Next, we discuss the results of
Hybrid Monte-Carlo (HMC) calculation
without diagonalized gauge fixing.
We generate the configurations by HMC and optimize
the variational parameters by the stochastic gradient descent method.
Specifically we use ADADELTA~\cite{zeiler2012adadelta} as an optimizer
to minimize the cost function evaluated using the generated configurations.
Figure \ref{Fig:expect} shows APF and
the expectation value of observables as a function of $\mu$.
After some steps of optimization,
$1-|\langle e^{i\theta}\rangle|$ becomes 3 to 100 times smaller
than before.
So we can weaken the sign problem in this theory.
The expectation values are compared with the exact solutions,
and results agree with exact results within the error.

\begin{figure}[bthp]
\begin{center}
\includegraphics[width=0.5\textwidth]{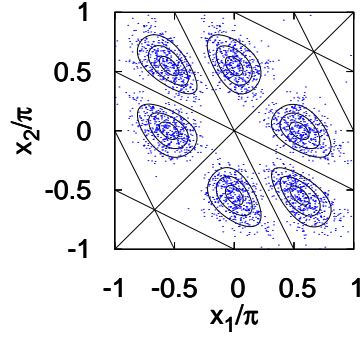}%
\end{center}
\caption{
The scatter plot of eigenvalue of the link variable $\mathcal{U}$.
White curves represents the contour of $|Je^{-S}|$
with diagonalized gauge fixing results.
The results are calculated at $\mu/T=1.0$.
}
\label{Fig:MCdist}
\end{figure}

Finally,
in Fig.~\ref{Fig:MCdist} we compare the eigenvalue distribution
of the results from two different ways of complexifications.
We diagonalize the link variables $\mathcal{U}$ generated by HMC
without diagonalized gauge fixing,
\begin{align}
  \mathcal{U} = P~\mathrm{diag}(e^{iz_1},e^{iz_2},e^{-i(z_1+z_2)}) P^{-1},
\end{align}
where $P$ is a regular matrix.
As a result, the real part of $z_1$ and $z_2$ are distributed
in six separated regions.
This distribution is consistent with that with
diagonalized gauge fixing which is shown by the contour.
Thus the separated regions are found to be connected by the continuous
transformation and do not cause the multimodal problem.

\section{1+1D QCD}

Let us now proceed to discuss 1+1 dimensional QCD.
We use one species of staggered fermion without rooting
and take the strong coupling limit.
We take the Polyakov gauge without diagonalized gauge fixing,
where $U_{0,x} = I$ for $x_\tau \neq N_\tau$
with $I$ being the identity matrix,
and we complexify the link variables as in 0+1 dimensional QCD
but in a more symmetric form as,
\begin{align}
  \mathcal{U}_{\mu, x} &=
  U_{\mu,x} \frac{H_{\mu, x}(U)}{\sqrt[3]{H_{\mu, x}(U)}}, \\
  H_{\mu, x}(U) &=
  I + \sum_{i=1}^8 \lambda_i y_{\mu, x, i}(\mathrm{Re}~U, \mathrm{Im}~U),
\end{align}
except for $\mu=0,~x_\tau \neq N_\tau$.
Here $y_{\mu, x, i}$ are the outputs of neural network,
and inputs of neural network are the real and imaginary part of
the link variables.

\begin{figure}[bthp]
\begin{center}
\includegraphics[width=0.45\textwidth]{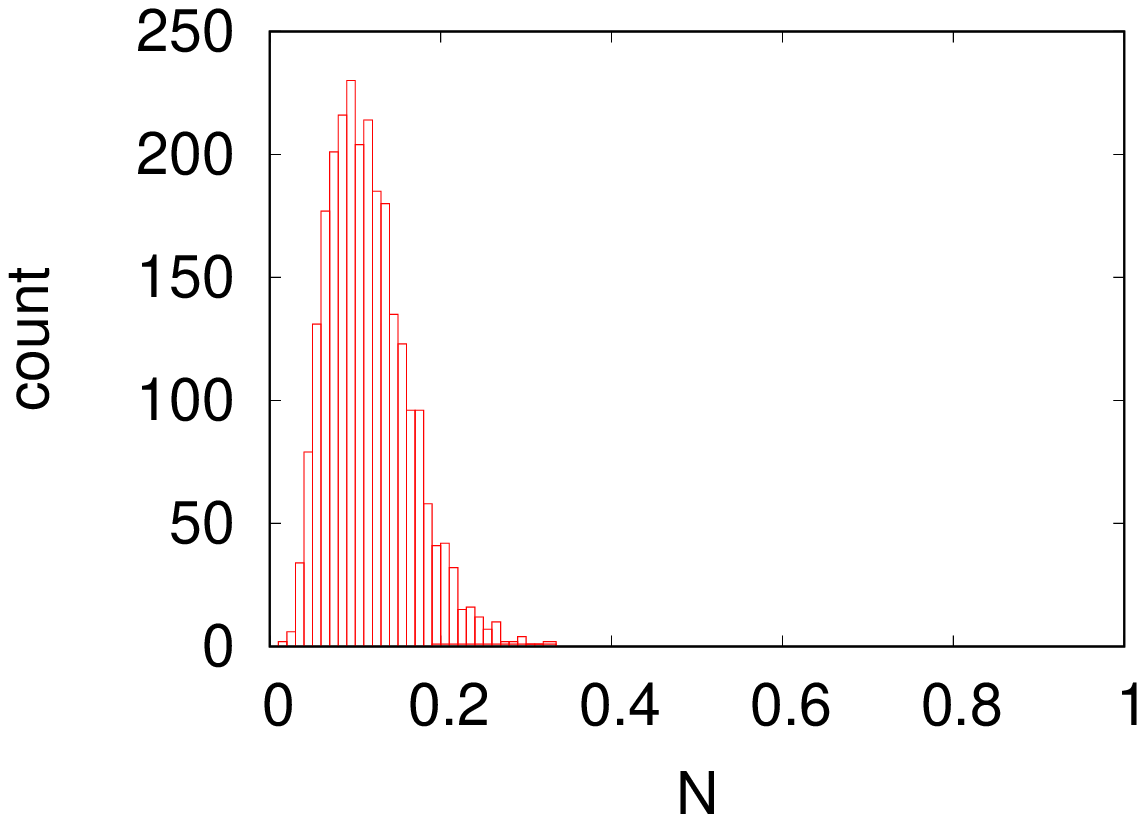}
\includegraphics[width=0.45\textwidth]{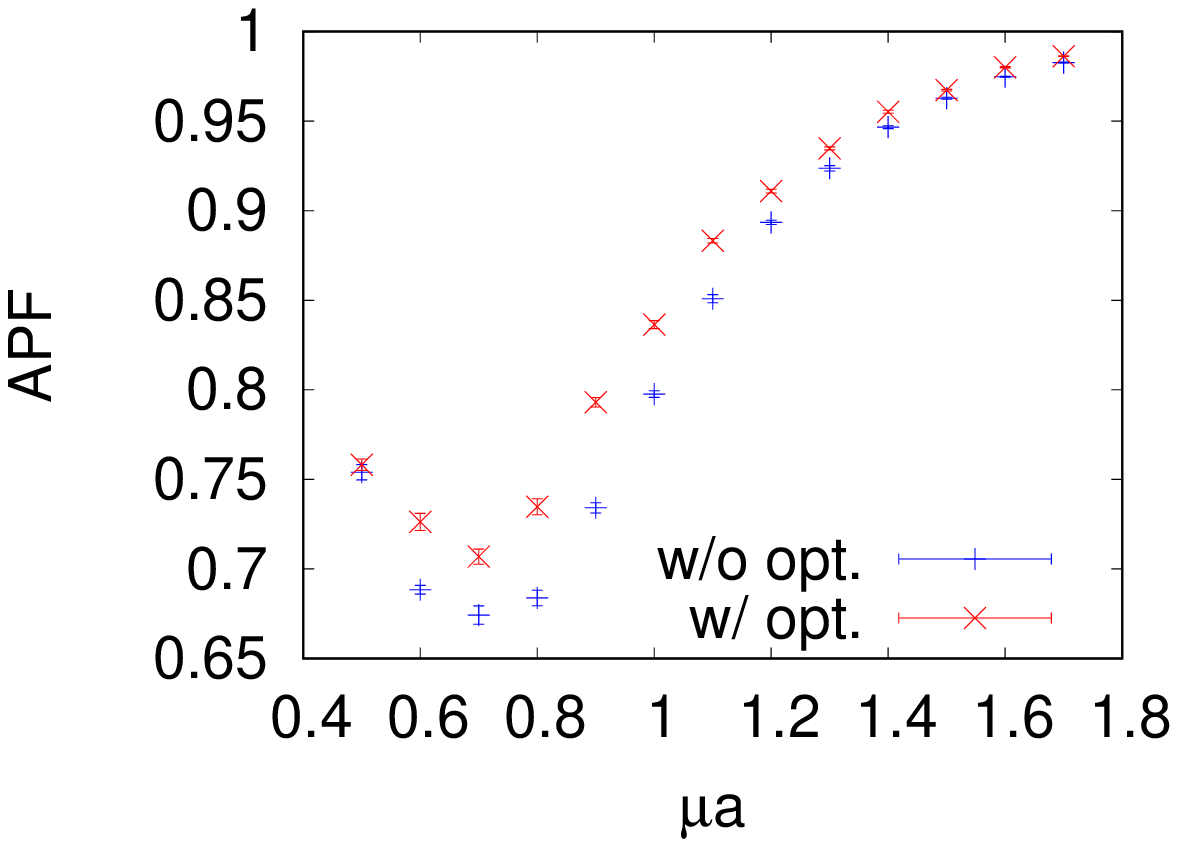}
\end{center}
\caption{
The unitary norm at $\mu a = 0.7$ (left),
and the average phase factor with and without optimization (right).
}
\label{Fig:2DQCD}
\end{figure}

We show the results on the $2^2$ lattice
and the bare mass is taken to be 0.1.
The left panel of Fig.~\ref{Fig:2DQCD}
shows the histogram of the unitary norm defined as,
\begin{align}
  N = \max_{\mu,x}\mathrm{tr}
    (\mathcal{U}^{\dagger}_{\mu,x}\mathcal{U}_{\mu,x}-1)^2.
\end{align}
The finite unitary norm means
the some deformations from $SU(3)$ are caused by the optimization.

The right panel of Fig.~\ref{Fig:2DQCD}
shows APF.
Original APF of this theory is smaller than that in 0+1D QCD,
and have dip structure at finite chemical potential.
By the optimization, APF can be enhanced.
But the increase is not significant, and then
the upper bound of APF may exist.
In order to confirm the latter point,
we need to compare the results with that of Lefschetz thimble method
and to clarify the reasons of the existence of the
upper bound in the future work.

\section{Summary}

In the path optimization method,
we can regard the sign problem as an optimization problem.
Neural network and optimization methods developed
in machine learning are helpful.
We apply this method to 0+1D QCD with and without diagonalized gauge fixing.
Average phase factor becomes large and exact results are reproduced
in observable calculations.
Without gauge fixing, eigenvalue distribution agrees with gauge fixed results,
and the hybrid Monte-Carlo sampling has been demonstrated
to work well, while the statistical weight distribution has
six separated regions and multimodal problem can be expected.
In 1+1D QCD, the average phase factor is
found to increase by the optimization, but we have not yet found the path
having a large average phase factor, e.g. larger than $0.9$ in some regions
of the chemical potential.


\section*{Acknowledgments}
This work is supported in part by
the Grants-in-Aid for Scientific Research from JSPS (Nos. 
18J21251, 
18K03618, 
19H01898, 
and
19H05151), 
and by the Yukawa International Program for Quark-hadron Sciences (YIPQS).

\bibliographystyle{JHEP}
\bibliography{ref}

\end{document}